\documentclass[aip,adv,reprint,floatfix,amsmath,amssymb]{revtex4-2}

\usepackage{graphicx}
\usepackage{dcolumn}
\usepackage{bm}
\usepackage{accents}
\usepackage[utf8]{inputenc}
\usepackage[T1]{fontenc}
\usepackage{mathptmx}
\usepackage{etoolbox}
\usepackage{textcomp,mathcomp}

\makeatletter
\def\@email#1#2{%
 \endgroup
 \patchcmd{\titleblock@produce}
  {\frontmatter@RRAPformat}
  {\frontmatter@RRAPformat{\produce@RRAP{*#1\href{mailto:#2}{#2}}}\frontmatter@RRAPformat}
  {}{}
}%
\makeatother
\begin{document}


\title{Automated CR-39 pit detection for energy and angular characterization of laser-driven ion beams}
\author{Wei Lin$^{1}$, Tzu-Yao Huang$^{1}$, Shih-Fan Yang$^{1}$, Shih-Wei Wang$^{1}$, Wu-Cheng Chiang$^{1}$, Kuo-Yu Hsiao$^{1}$, Chun-Han Chen$^{1}$, Yan-Syu Liu$^{1}$, Jun-Yi Chen$^{1}$, Ya-Po Yang$^{1}$, Chen-Kang Huang$^{1,2}$, Hsu-hsin Chu$^{1,2}$, and Chih-Hao Pai$^{1,2}$}
\email{chpai@phy.ncu.edu.tw}
\affiliation{$^1$Department of Physics, National Central University, Taoyuan 320317, Taiwan\\
    $^2$Center for High Energy and High Field Physics, National Central University, Taoyuan 320317, Taiwan}%

\date{July 28, 2026}

\begin{abstract}
    Accurate characterization of ion beams from sub-MeV to multi-MeV energies is essential for diagnostics in laser-driven ion acceleration and fusion research. CR-39 detectors are widely used to measure absolute ion yields, while ion energy and incidence angle can be inferred from etched-pit geometries. In this study, we present an automated CR-39 pit detection method that enables reliable extraction of pit geometries from two-dimensional optical micrographs, with improved discrimination of overlapping pits under high-flux conditions. This capability enables reliable analysis in high-density regimes where conventional methods are limited by significant track overlap. By integrating the detected pit shapes with a pit-formation model based on Monte Carlo simulations of restricted energy loss, energy and angular characterization of laser-driven proton beams is achieved with enhanced precision and computational efficiency. The proposed framework provides a robust and scalable approach for automated CR-39 analysis and is applicable to a broad range of ion-beam diagnostics and spectroscopy.
\end{abstract}

\maketitle

\section{\label{sec:Introduction}Introduction}
Columbia Resin-39 (CR-39) is a widely used solid-state nuclear track detector that records ion tracks through localized scission of polymer chains induced by energetic ions~\cite{cartwright1978nuclear}. These latent tracks are subsequently revealed by chemical etching, a process governed by both the etching conditions and the interaction between incident ions and the polymer chains. Owing to its chain-like molecular structure, CR-39 exhibits high sensitivity to light ions and has been widely applied in nuclear and particle physics~\cite{fleischer2022nuclear,durrani2013solid}, radiation therapy and dosimetry~\cite{kodaira2019contribution,castillo2013fast}, laser-driven ion acceleration~\cite{singh2022low,shou2025laser}, and laser-assisted fusion studies~\cite{giuffrida2020high,bonvalet2021energetic}.

The formation of latent tracks and their subsequent development into etch pits during chemical etching are determined by the initial characteristics of the incident ions, commonly modeled using the restricted energy loss (REL) criterion~\cite{fleischer1967criterion,benton1970latent}. This dependence manifests in the track etch rate, as demonstrated experimentally by Somogyi~\cite{somogyi1980development}. The resulting pit geometry is described by track-etching models originally developed by Fleischer and Price~\cite{fleischer1964glass} and later generalized by Somogyi and Szalay~\cite{somogyi1973track}.

Despite the capability of CR-39 detectors to provide detailed information on individual ions, their application in high-flux environments is often limited by the labor-intensive and time-consuming nature of pit identification and measurement. In laser-driven experiments, where large numbers of tracks are produced, efficient extraction of ion information from etched CR-39 samples requires automated and robust image-analysis methods. Conventional perimeter-tracking methods, such as the dynamic edge detection implemented by Fews~\cite{fews1992fully}, identify overlapping tracks by searching for reentrant corners along the contour. This approach facilitates precise morphological extraction for isolated pits; however, it is susceptible to failure in high-flux regions where severe geometric merging or image noise obscures these critical reentrant features. More recently, Seimetz \textit{et al.} utilized the ROOT framework for automated track identification within predefined diameter ranges~\cite{seimetz2018spectral}. This method accelerates the processing of large datasets; however, its assumption of circular pit openings fundamentally discards the incidence-angle information encoded in elliptical tracks.

Three-dimensional confocal microscopy enables full reconstruction of pit morphology, but at the cost of substantially increased data acquisition and processing complexity~\cite{vaginay2001}. Furthermore, the CR-39 response depends not only on etching conditions but also on manufacturing variations~\cite{lahmann2020cr}, necessitating careful calibration when relating pit geometry to ion parameters. Under high-flux conditions, commonly encountered in laser-driven ion acceleration and fusion experiments, the ability to resolve overlapping pits and to perform computationally efficient analysis is therefore essential.

In this study, we develop an automated detection algorithm for etched pits in two-dimensional optical micrographs of alkaline-etched CR-39, enabling accurate determination of pit size and particle number. The computational framework combines a global detection scheme, optimized for resolving overlapping pits, with local least-squares ellipse fitting to extract pit geometries with high precision. In contrast to conventional perimeter-tracking approaches that rely on reentrant-corner identification, the proposed framework remains effective in high-flux regions where severe pit overlap can obscure contour features. Furthermore, unlike circular-track-based identification methods, the ellipse-fitting procedure preserves the incidence-angle information encoded in the pit geometry, enabling simultaneous reconstruction of particle energy and trajectory. The implementation supports efficient GPU acceleration and maintains robustness under high-yield conditions with substantial pit overlap. To establish the relationship between pit geometry and ion parameters, calibration experiments were performed using laser-driven multi-MeV proton beams generated via the target normal sheath acceleration (TNSA) mechanism at the 100-TW laser facility at National Central University. The extracted pit dimensions are compared with a theoretical pit-formation model based on Monte Carlo simulations of restricted energy loss, enabling reconstruction of ion energy and incidence angle from the measured major and minor axes of the fitted ellipses.

\section{\label{sec:Setup}Experimental Setup}
\begin{figure}[tbp]
    \centering
    \includegraphics[width=\columnwidth]{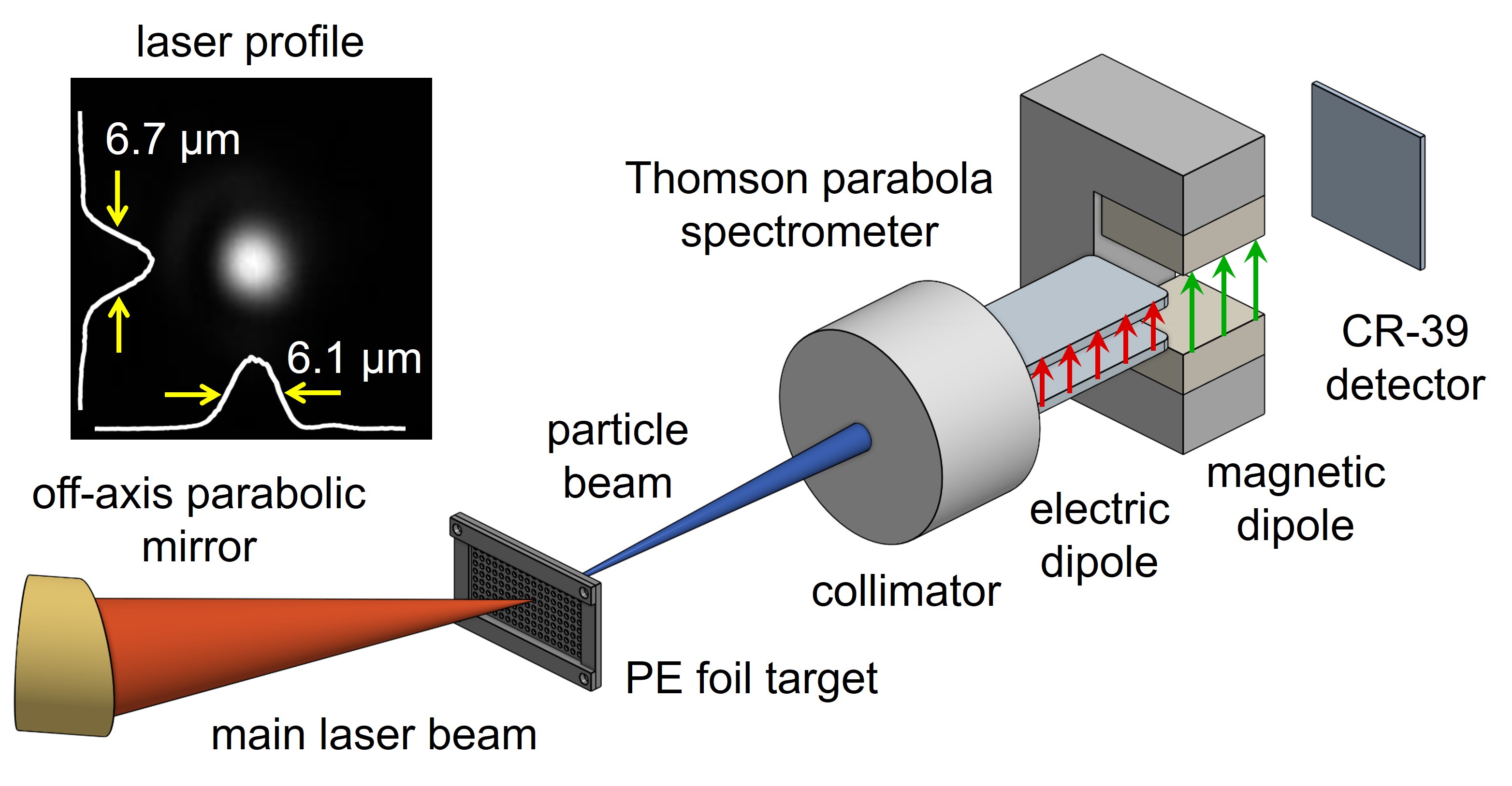}
    \caption{\label{Fig1:setup} Experimental setup for validation of the pit-detection algorithm and ion-parameter reconstruction. \emph{Inset}: focal spot image and corresponding intensity profile. The PE foil target is mounted on a holder capable of withstanding $\sim$150 laser shots. The downstream Thomson parabola spectrometer collects ions within a divergence angle of $\sim$1~mrad and energies above 0.3~MeV. The red and green arrows indicate the directions of the electric and magnetic fields of the TPS, respectively.}
\end{figure}

The experiment was performed using a 100-TW Ti:sapphire laser system based on chirped-pulse amplification at National Central University~\cite{Hung2014APB}. The schematic of the experimental setup is shown in Fig.~\ref{Fig1:setup}. An 800-nm, $p$-polarized laser pulse is focused onto the target using an f/6 off-axis parabolic mirror (OAP), producing a focal spot with a full width at half maximum (FWHM) diameter of 6.4~$\mu $m and 81.6\% energy enclosed within a Gaussian-fit profile. The laser pulse duration is 43~fs (FWHM), with a temporal contrast ratio of $10^{-9}$ at 250~ps before the main pulse. The corresponding peak intensity is $5.0\times10^{19}$~W/cm$^2$. A polyethylene (PE) foil with a thickness of 11~$\mu $m is used as the target and positioned at the focal plane.

Ion beams generated via the TNSA mechanism were characterized using a Thomson parabola spectrometer (TPS) aligned with the target normal. The customized TPS consists of a collimator positioned 30~cm from the target, followed by an electric dipole and a magnetic dipole with field strengths of 200~kV/m and 0.31~T, respectively, enabling charge-to-mass-dependent deflection of ions. Downstream CR-39 detectors record the deflected ion traces, from which the energy spectra are reconstructed. The ion energies measured by the TPS also serve as inputs for subsequent pit-formation modeling (Sec.~\ref{se:result}).

Chemical etching of CR-39 samples was performed in a 7.25~M NaOH solution at $(80\pm1)$~\textdegree C for 30 min under continuous stirring to ensure uniform etching conditions. Based on the restricted energy loss model and the resulting pit-size distributions (Sec.~\ref{se:result}), an etching duration of 0.5~h is identified as optimal for maximizing the diagnostic dynamic range. At this duration, shallow tracks generated by high-energy protons exceed the optical diffraction limit, while severe pit overlap from low-energy protons in high-flux regions is largely avoided. Optical micrographs of the etched samples were acquired using a TASLImage$^\mathrm{TM}$ microscope system~\cite{taslimage}.

\section{\label{sec:Algorithm}PIT DETECTION ALGORITHM ON ETCHED CR-39 FOR ION SPECTRAL ANALYSIS}
\subsection{\label{sec:model}Fundamentals of numerical modeling}

\begin{figure}[tbp]
    \centering
    \includegraphics[width=0.8\columnwidth]{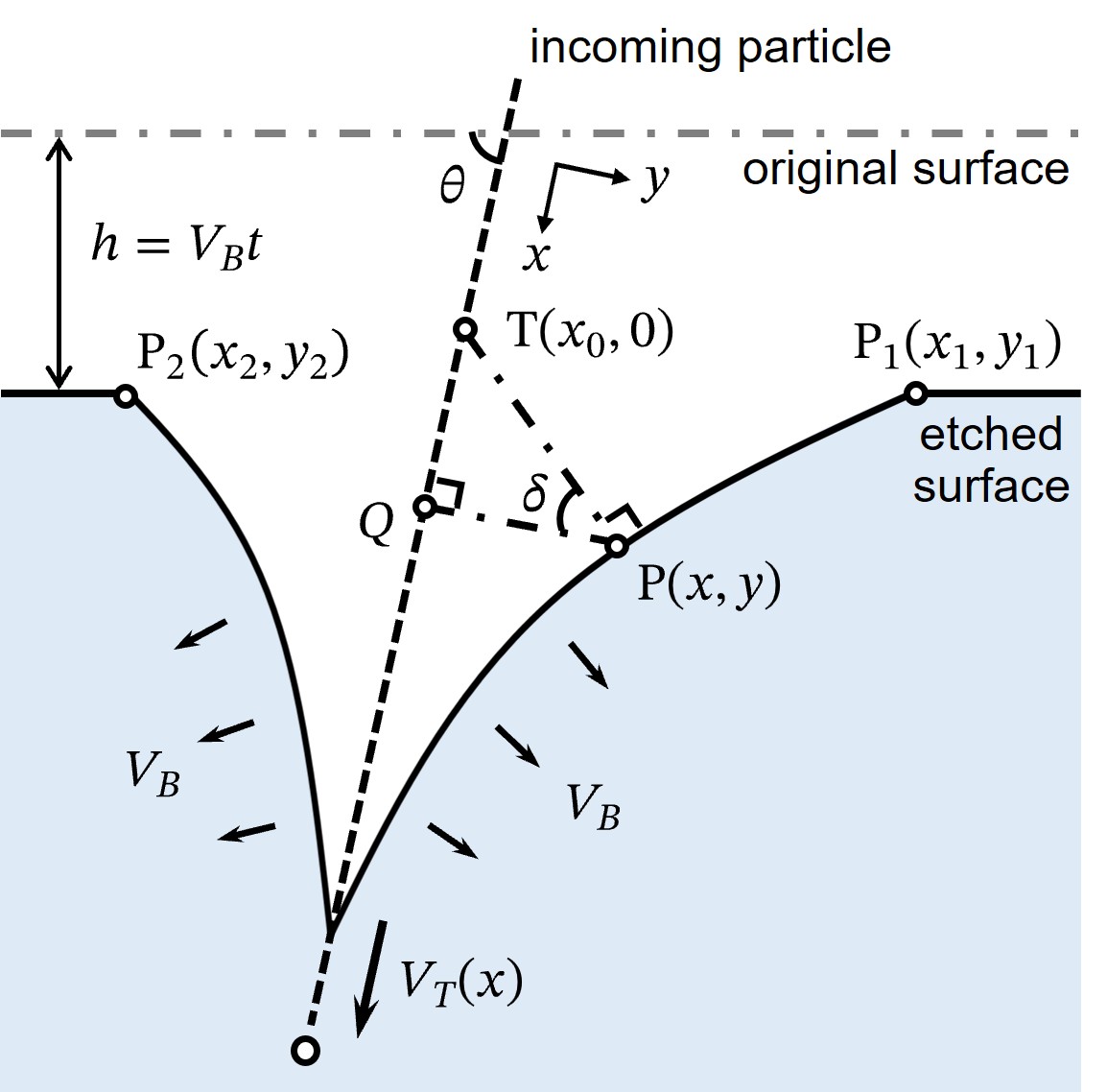}
    \caption{\label{Fig2:schematic} Schematic of pit formation during chemical etching. The coordinate $x$ denotes the track axis along the ion trajectory. After a surface removal depth $h_B$, the pit walls expand laterally at the bulk etch rate $V_B$ and advance along the track at the track etch rate $V_T(x)$ until reaching the terminal point.}
\end{figure}

Pit formation in etched CR-39 results from the difference between the etch rate along the ion-damaged track, $V_T$, and
that of the undisturbed bulk material, $V_B$, as schematically illustrated in Fig.~\ref{Fig2:schematic}. For a homogeneous substrate, $V_B$ is constant, whereas $V_T$ varies along the ion track coordinate $x$, governed by the local damage density characterized by the
restricted energy loss (REL). Previous studies~\cite{somogyi1980development} have shown that the etch-rate ratio, defined as $V \equiv V_T/V_B$, can be expressed as
\begin{eqnarray}\label{eq:ratioREL}
    V(\text{REL})=1+\alpha\cdot \text{REL}^{\beta},
    \label{eq:one}
\end{eqnarray}
where $\alpha$ and $\beta$ are fitting parameters that depend on the polymer composition, fabrication process, and etching conditions.

The geometrical model of pit formation, originally developed by Somogyi and Szalay~\cite{somogyi1973track}, describes the evolution of the pit walls as a function of the etch-rate ratio $V$, the surface removal depth $h_B\equiv V_B t$, and the ion incidence angle $\theta$, where $t$ denotes the etching time. As illustrated in Fig.~\ref{Fig2:schematic}, an arbitrary surface point $P(x,y)$ can be traced back to a source point $T(x_0,0)$ along the ion trajectory. The etching time can be expressed as

\begin{eqnarray}
    t=\frac{h_B}{V_B}=\int_{0}^{x_0}\frac{1}{V_T(x)}dx+\frac{y}{V_B\cos(\delta)},
    \label{eq:two}
\end{eqnarray}

\noindent where $\delta = \arcsin{\left[V^{-1}(x_0)\right]}$ (see the supplementary material, Sec.~I.A), and the trajectory satisfies $x = x_0 + y \tan \delta$ from geometrical considerations.

Defining $h(x_0) \equiv \int_0^{x_0} V^{-1}(x)\, dx$, the pit-wall profile can be parameterized as

\begin{subequations}
    \label{eq:para-walls}
    \begin{align}
        x(x_0) & = x_0 + \left[ h_B - h(x_0) \right]/V(x_0),
        \label{subeq:para-walls-1}                                      \\
        y(x_0) & = \left[ h_B - h(x_0) \right]/V(x_0)\sqrt{V^2(x_0)-1}.
        \label{subeq:para-walls-2}
    \end{align}
\end{subequations}

The pit opening is determined by the edge points $P_1(x_1,y_1)$ and $P_2(x_2,y_2)$, which satisfy

\begin{subequations}
    \label{eq:p1p2conds}
    \begin{align}
        h_B = x(x_{01}) \sin\theta + y(x_{01}) \cos\theta,
        \label{eq:p1p2conds-1} \\
        h_B = x(x_{02}) \sin\theta - y(x_{02}) \cos\theta.
        \label{eq:p1p2conds-2}
    \end{align}
\end{subequations}
where $x_{01}$ and $x_{02}$ denote the source points associated with $P_1$ and $P_2$, respectively. While $V_T$ remains nearly constant over the distance $|x_{01} - x_{02}|$, the pit opening can be approximated by an ellipse (see the supplementary material, Sec.~I.B), with major and minor axes given by

\begin{subequations}
    \label{eq:ellipse-dims}
    \begin{align}
        a & = (y_1 + y_2)/\sin\theta,
        \label{subeq:ellipse-dims-1}  \\
        b & = 2\sqrt{y_1 y_2}.
        \label{subeq:ellipse-dims-2}
    \end{align}
\end{subequations}

The evolution of pit geometry for a given ion characteristics is obtained by fitting Eqs.~\eqref{subeq:ellipse-dims-1} and \eqref{subeq:ellipse-dims-2} to the measured major and minor axes, $a$ and $b$, for different ion energies at a known incidence angle. The required REL data for each ion species are calculated using the Geant4 Monte Carlo framework.

\subsection{Procedures for pit detection\label{sse:pitdet}}

\begin{figure}[tbp]
    \centering
    \includegraphics[width=0.8\columnwidth]{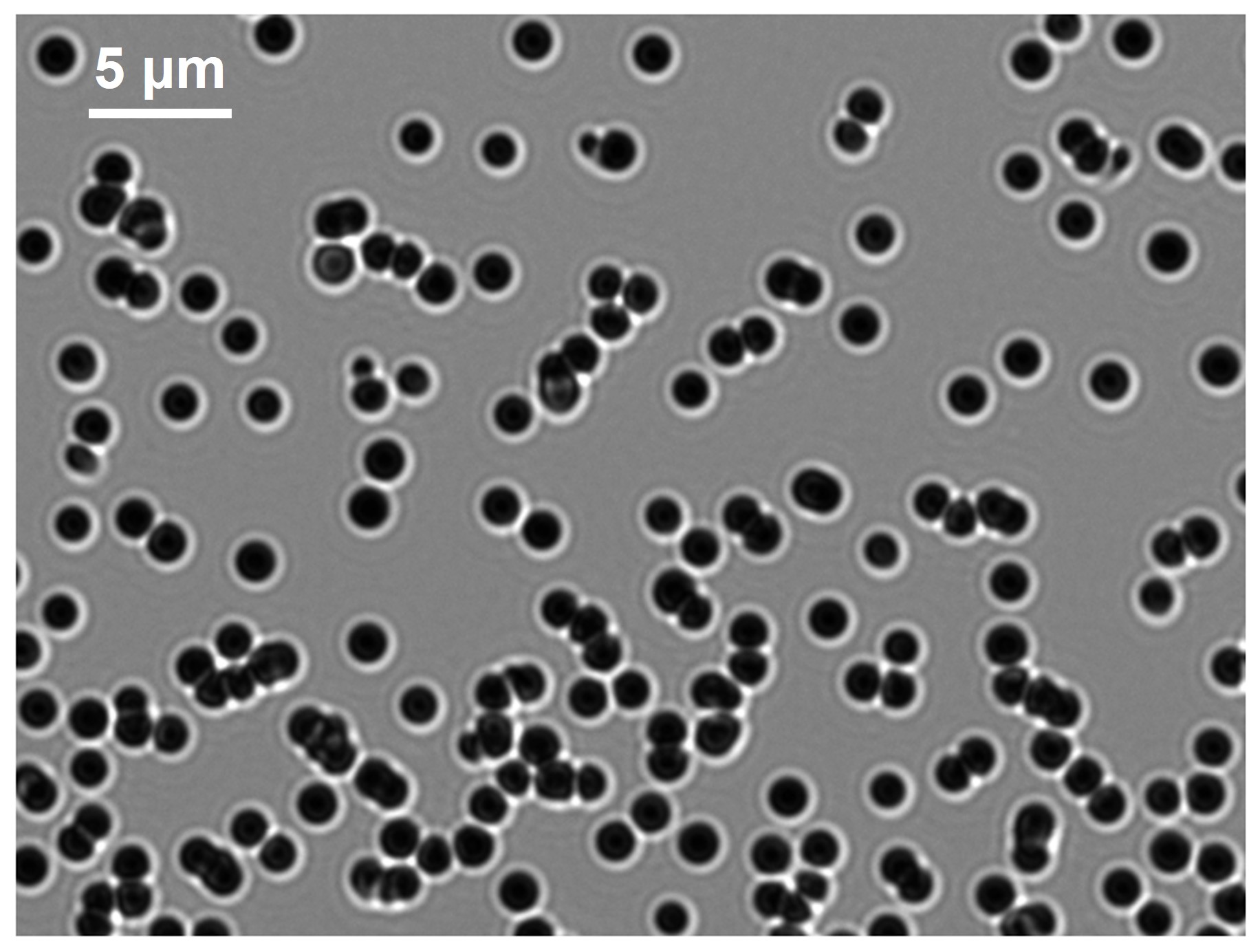}
    \caption{\label{fig:cr39-sample} Micrograph of a CR-39 detector following laser-driven proton irradiation and a 0.5-h etching period, illustrating multiple overlapping etch pits.}
\end{figure}

Given the relation between pit openings and ion parameters, precise measurement of pit opening shape allows for the retrieval of spectrometric information for different ion species. To that end, 8-bit grayscale images of the proton-irradiated CR-39 surface are acquired using the TASLImage\texttrademark{} microscope system \cite{taslimage}, as shown in Fig.~\ref{fig:cr39-sample}. The image processing framework is divided into two stages: global peak detection and local shape identification. This modular structure ensures both high detection accuracy and computational efficiency.

\subsubsection{Global peak detection\label{ssse:peakdet}}

\begin{figure}[tbp]
    \centering
    \includegraphics[width=\columnwidth]{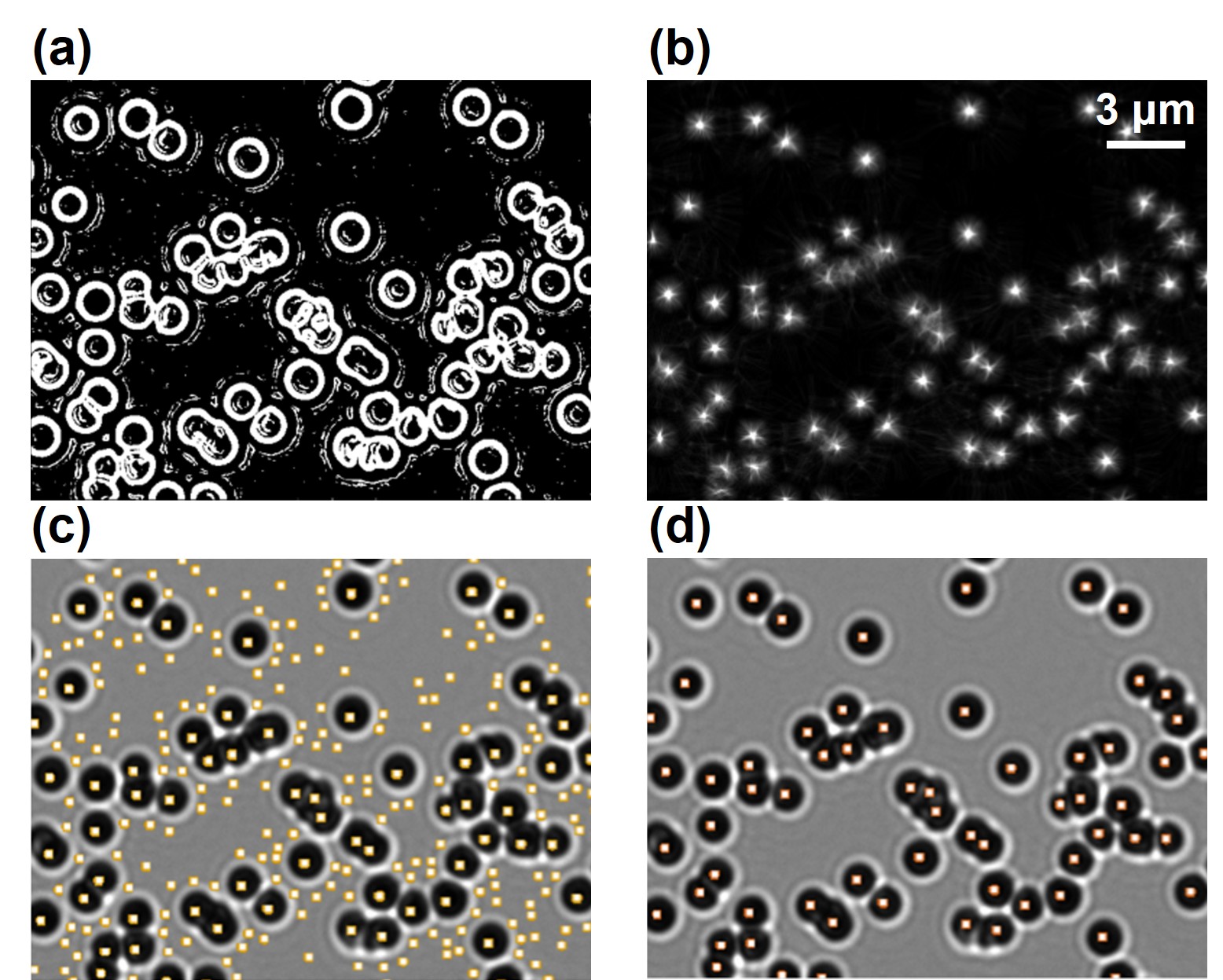}
    \caption{\label{fig:peakdet}Peak-identification for CR-39 images. (a) Edge mask from Otsu global thresholding applied to the gray-level gradient. (b) Accumulator space produced by Hough-transform voting. (c) Initial peak-detection with candidate peaks marked with yellow dots. (d) Final candidate pits after Gaussian smoothing and maximum filtering.}
\end{figure}

In the stage of global peak detection, cone tips of ion tracks are identified through analysis of grayscale intensity gradients, as illustrated in Fig.~\ref{fig:peakdet}(a). Edge points are extracted using Otsu's adaptive thresholding method~\cite{otsu1975threshold}, and employed to construct straight-line segments through a GPU-accelerated implementation of Bresenham's algorithm~\cite{bresenham1998algorithm}. A Hough-transform-based voting scheme~\cite{hough1962method} is subsequently applied to determine candidate pit centers from the spatial density of line intersections, as shown in Fig.~\ref{fig:peakdet}(b). The peak position is defined as the location of maximal line accumulation in the parameter space.

As illustrated in Fig.~\ref{fig:peakdet}(c), this peak-detection approach is susceptible to contamination from background noise and diffraction patterns arising from the pit boundaries. To mitigate these effects, a maximum filter is applied to the smoothed parameter space, with the kernel size chosen according to a user-specified minimum spacing between adjacent pits. The refinement, shown in Fig.~\ref{fig:peakdet}(d), enhances robustness to variability in edge-point detection and effectively suppresses unwanted features arising from surface scratches and particulate contamination.

\subsubsection{Local ellipse detection\label{ssse:ellidet}}

The major and minor axes of the elliptical pit openings are obtained via localized fitting of individual pits. For each detected peak, multiple analysis windows centered on the peak are evaluated. The direct least-squares ellipse fit of Fitzgibbon \textit{et al.}~\cite{fitzgibbon1999direct} (details in the supplementary material, Sec.~II) is computed in parallel on GPUs. A conic section is represented in matrix form with the parameter vector $\mathbf{p} = [p_1 \;\; p_2 \;\; p_3 \;\; p_4 \;\; p_5 \;\; p_6]^T$ ($T$ is the matrix transpose operation), and the data matrix has the $i$th row $\mathbf{D}_i = [x_i^2 \;\; x_i y_i \;\; y_i^2 \;\; x_i \;\; y_i \;\; 1]$. To ensure the fitted conic strictly represents an ellipse rather than a hyperbola or parabola, an equality constraint ($4p_1p_3 - p_2^2 = 1$) is introduced to guarantee a unique elliptical solution. The ellipse constraint is represented by a $6\times6$ matrix $\mathbf{C}$, where $\mathbf{C}_{1,3} = \mathbf{C}_{3,1} = 2$, $\mathbf{C}_{2,2} = -1$, and all other entries are zero. A Lagrangian formulation reduces the constrained least-squares problem to a generalized eigenvalue problem by setting the corresponding derivatives to zero

\begin{equation} \label{eq:eprob1}
    \mathbf{S} \mathbf{p} = \lambda \mathbf{C} \mathbf{p},
\end{equation}
where $\mathbf{S} = \mathbf{D}^T \mathbf{D}$ and $\lambda$ is the Lagrangian multiplier. The solution vector $\mathbf{p'} = [p'_1 \;\; p'_2 \;\; p'_3 \;\; p'_4 \;\; p'_5 \;\; p'_6]^T$ leads to the expression for the ellipse center $(x_c, y_c)$, given by

\begin{equation} \label{eq:center}
    x_{c} = \frac{p'_2 p'_5 - 2p'_3 p'_4}{{p'_2}^2-4p'_1 p'_3}, \quad y_{c} = \frac{p'_2 p'_4 - 2p'_1 p'_5}{{p'_2}^2-4p'_1 p'_3}.
\end{equation}

The precise determination of the major and minor axes relies on reducing numerical instability. To this end, the coordinate system is centered using $x' = x - x_c$ and $y' = y - y_c$, and the ellipse equation is reformulated into the form

\begin{equation} \label{eq:reform}
    \tilde{p}_1 {x'}^2 + \tilde{p}_2 x' y' + \tilde{p}_3 {y'}^2 = 1,
\end{equation}
where $\tilde{p}_1, \tilde{p}_2, \text{ and } \tilde{p}_3$ are the normalized coefficients after eliminating the linear terms through centering. The eigenvalues of the corresponding matrix representation of the quadratic form are determined by
\begin{equation} \label{eq:eigenvalues}
    \lambda_{\min,\max} = \frac{1}{2} \left[( \tilde{p}_1 + \tilde{p}_3) \mp \sqrt{(\tilde{p}_1 - \tilde{p}_3)^2 + \tilde{p}_2^2} \right],
\end{equation}
which allows the measured major and minor axes, $a_m$ and $b_m$, to be computed from the relations
\begin{equation} \label{eq:ab}
    a_m = \frac{1}{\sqrt{\lambda_{\min}}}, \quad b_m = \frac{1}{\sqrt{\lambda_{\max}}}.
\end{equation}
The inclination of the ellipse, which gives the information of incident angle, is determined by $\theta = \arctan\left( a_m/b_m \right)$.

\begin{figure}[tbp]
    \centering
    \includegraphics[width=0.8\columnwidth]{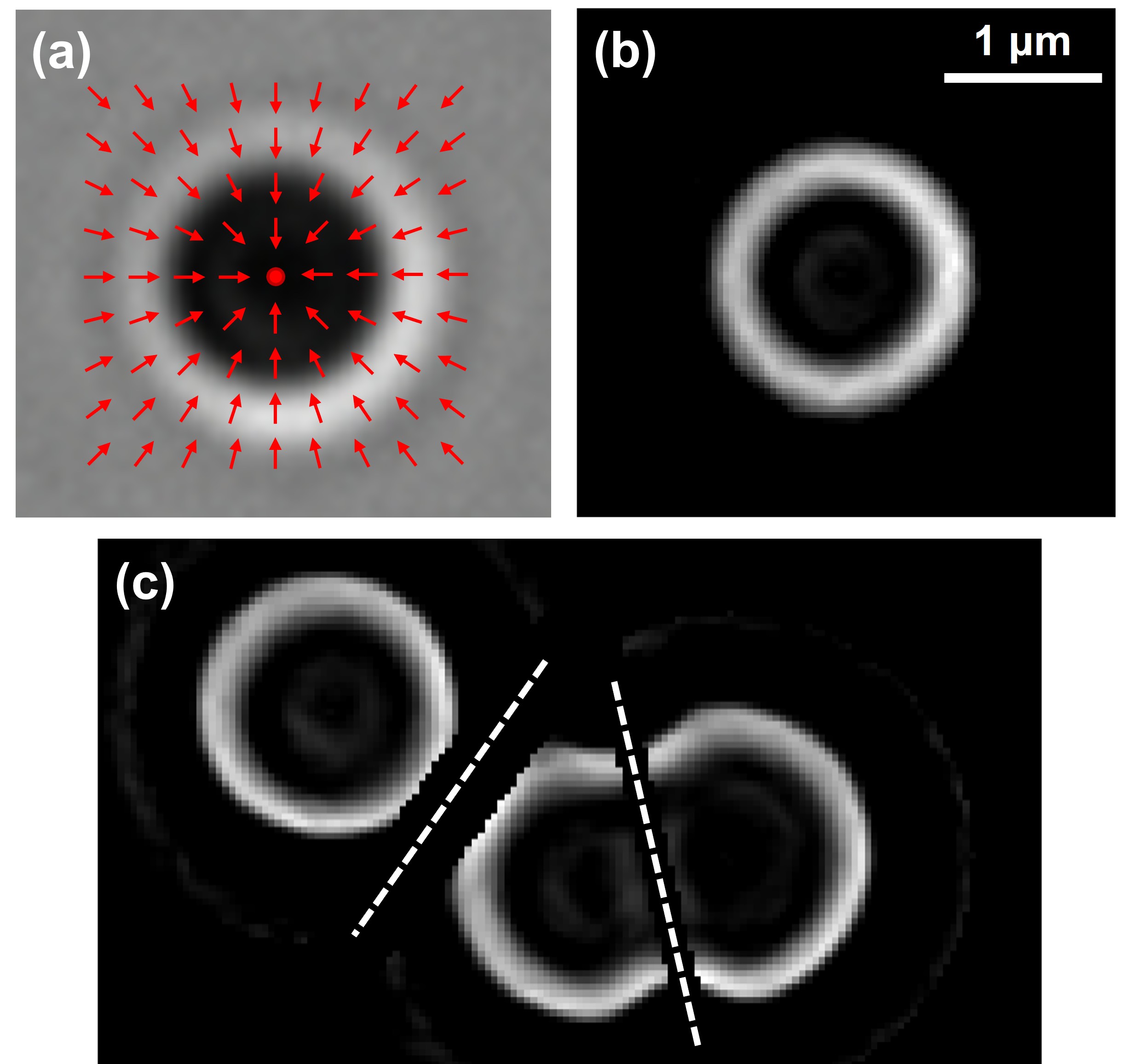}
    \caption{\label{fig:ellidet}Ellipse-fitting and overlap-separation procedure for CR-39 pits. (a) Pit micrograph with the unit radial vector field from Eq.~\eqref{eq:vecfield} overlaid. (b) Dot product map between the grayscale gradient and the unit radial field. (c) Separation of overlapping pits to minimize mutual interference between adjacent detections.}
\end{figure}

The determination of fitting points is based on the gradient similarity to a reference pit, characterized by a unit vector field (Fig.~\ref{fig:ellidet}(a))

\begin{equation} \label{eq:vecfield}
    \mathbf{V}(x, y) = \frac{(x - x_c,\, y - y_c)}{\sqrt{(x - x_c)^2 + (y - y_c)^2}}.
\end{equation}

The dot product between the grayscale gradient and $\mathbf{V}(x, y)$ provides a scalar score at each point, as shown in Fig.~\ref{fig:ellidet}(b). A subsequent Otsu thresholding is then applied to select pixels with both large gradient magnitudes and strong directional agreement, thereby systematically decoupling the pit contours from optical diffraction fringes and surface defects.

In regions of high track density, overlapping pits produce merged elliptical profiles with indistinct boundaries, complicating the discrimination of individual openings. To address this issue, a separation line is placed in the middle between each pair of overlapping pit peaks to separate the detection areas (Fig.~\ref{fig:ellidet}(c)). A user-defined exclusion region around each separation line is omitted from the detection process, which breaks the coupling between intersecting tracks and prevents bias from geometrical fitting. While the exclusion-region width must be chosen with care, the procedure remains robust and can reliably fit ellipses using only partial segments when pit overlap occurs.

\section{Results\label{se:result}}

\begin{figure}[tbp]
    \centering
    \includegraphics[width=0.7\columnwidth]{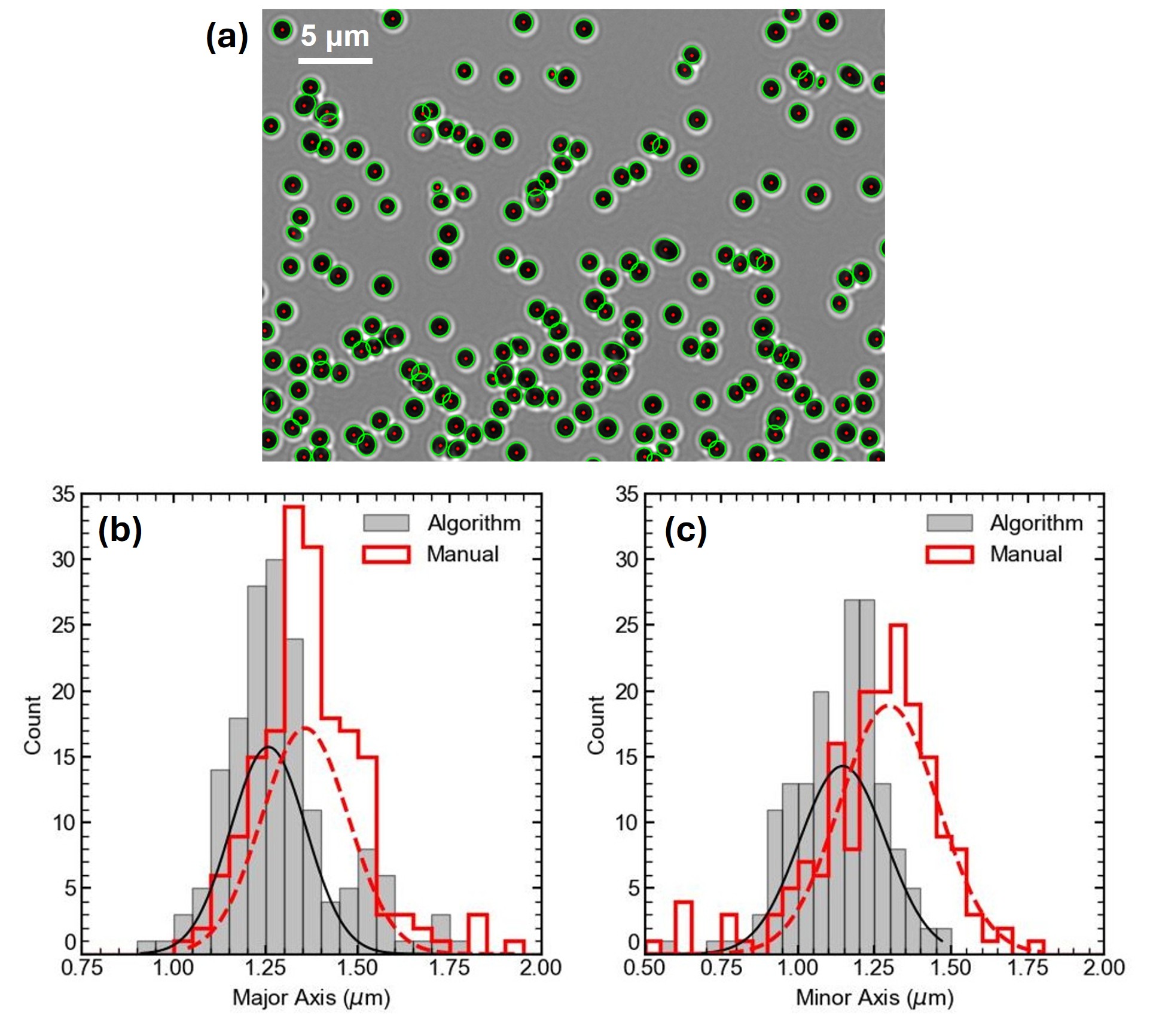}
    \caption{\label{fig:detres}Pit detection results. (a) Detected pits are outlined by green ellipses, with centers indicated by red dots. (b) Distributions of the major axes and (c) minor axes of the detected pits, comparing results obtained from the automated algorithm (gray histogram, solid-line fit) and manual measurements (red histogram, dashed-line fit). The comparison reveals a $\sim$10\% systematic geometric offset resulting from differences in boundary definition, with manual measurements relying on subjective visual identification of the pit edge and the automated algorithm defining the boundary at the location of the maximum grayscale gradient.}
\end{figure}

With appropriate parameter settings, the detection algorithm reliably extracts the pit positions and the corresponding major and minor axes, as shown in Fig.~\ref{fig:detres}(a). These data enable statistical analyses of pit-size distributions, divergence-angle distributions, and spatial profiles of the ion number density.

To evaluate the algorithm's performance, the major and minor axes of the same pit dataset were independently measured by manual tracing. A comparison between the automated and manual results, shown in Figs. 6(b) and 6(c), reveals a systematic difference of $\sim$10\% in the mean pit size. This discrepancy originates from the different definitions of the pit boundary. In manual tracing, the boundary is visually assigned to the outermost optical diffraction fringe, whereas the algorithm objectively defines it at the location of the maximum grayscale gradient. Because the pit-formation model is subsequently calibrated empirically, this geometric offset is fully incorporated into the calibration procedure and therefore does not affect the final energy reconstruction.

To further assess the uncertainty associated with manual measurements, the pits were independently analyzed by three researchers. The resulting mean pit sizes exhibited standard deviations of up to 9.4\%, primarily due to subjective differences in boundary identification, particularly in regions containing overlapping pits.

In contrast, the automated algorithm is fully reproducible and eliminates the random variability inherent in manual measurements. Its high computational efficiency further enables rapid processing of large datasets, providing consistent and objective results without operator-dependent bias.

\begin{figure}[tbp]
    \centering
    \includegraphics[width=0.7\columnwidth]{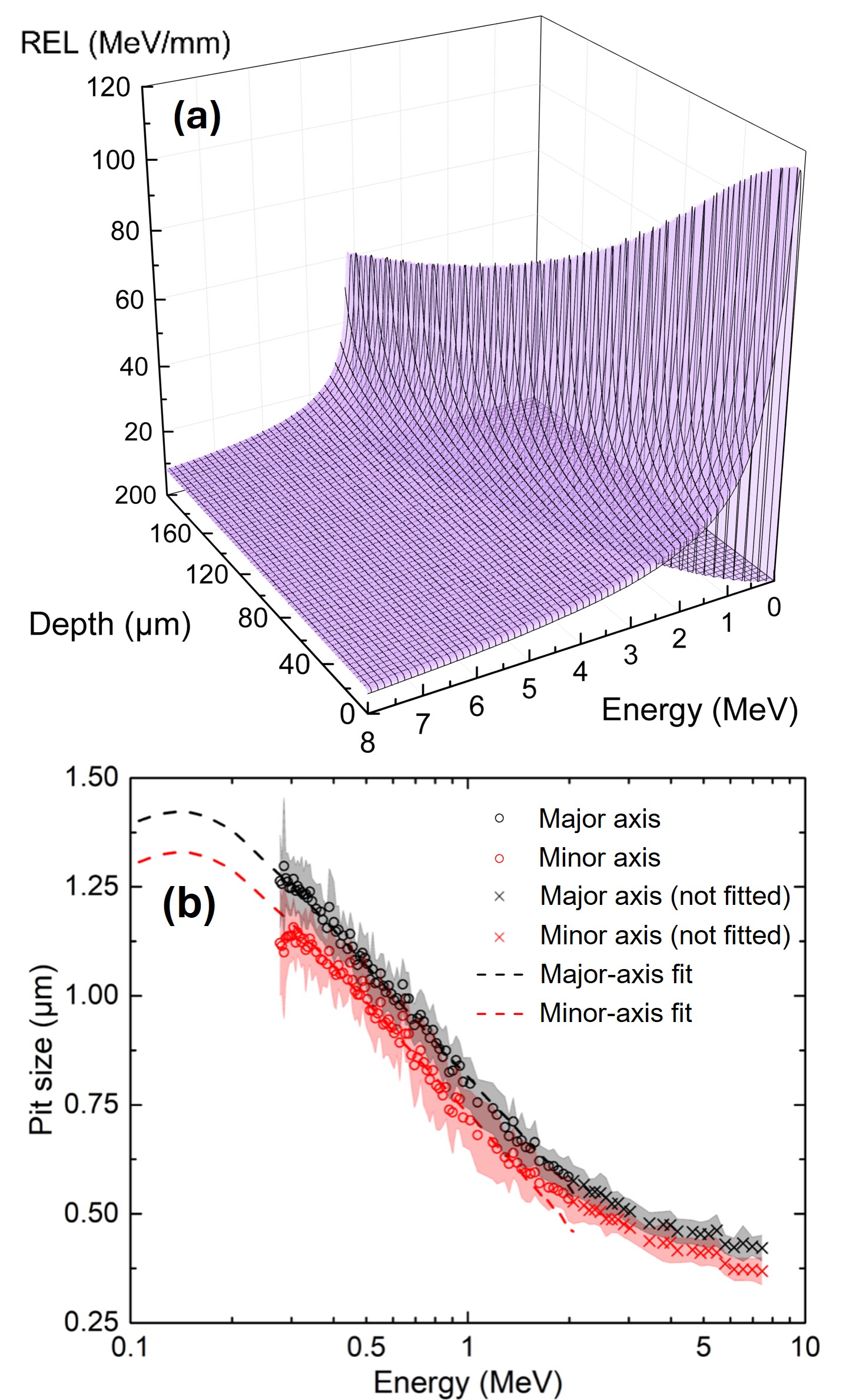}
    \caption{\label{fig:fitting}Calibration of the pit-formation model. (a) Restricted energy loss (REL) of protons with energies up to 8~MeV obtained from Geant4 simulations. (b) Fitting of the relationship between the major and minor axes and the proton energy at a fixed incident angle. Data points corresponding to energies above 2~MeV are excluded from the fitting procedure due to diffraction effects at the pit edges, which introduce significant deviations in the measured pit openings.}
\end{figure}

Subsequent analysis of ion characteristics is based on the pit-formation model with parameters $(\alpha, \beta)$ in Eq.~\eqref{eq:ratioREL}. These parameters are obtained by fitting the experimentally measured pit axes $(a_m, b_m)$, described in Sec.~\ref{sse:pitdet}, at different ion energies for a fixed angle of incidence. As a demonstration, a normally incident proton beam with energies from 0.3 to 7.4~MeV was characterized using CR-39s as detectors in the Thomson parabola spectrometer (Sec.~\ref{sec:Setup}). Etching was performed for $t=0.5$~h with a bulk etch rate of $V_B = 1.684 \pm 0.004~\mu\mathrm{m/h}$. The value of $V_B$ was determined from the mass loss of blank CR-39 samples etched for different durations and converted to an equivalent thickness removal using the detector surface area and material density. The REL along ion trajectories in CR-39 is simulated using the Geant4 Monte Carlo framework~\cite{agostinelli2003geant4}, as shown in Fig.~\ref{fig:fitting}(a). The measured pit sizes are fitted using Eqs.~\eqref{eq:ratioREL}--\eqref{eq:ellipse-dims}, as shown in Fig.~\ref{fig:fitting}(b), where the shaded regions represent measurement uncertainty. Data points above $\sim$2~MeV are excluded from the fitting because the corresponding pit dimensions approach the diffraction limit of the optical microscopy system, resulting in systematic deviations in both the major and minor axes. These deviations can be mitigated by extending the etching time if necessary. The resulting fitted parameters, $(\alpha, \beta) = [(1.007 \pm 0.002)\times 10^{-2},\, 1.501 \pm 0.001]$, have relative uncertainties below 0.2\% and show good agreement between the pit-formation model and the measured pit openings. These parameters are independent of the etching duration, which can therefore be varied across different regions of interest while maintaining consistent calibration.

\begin{figure}[tbp]
    \centering
    \includegraphics[width=0.8\columnwidth]{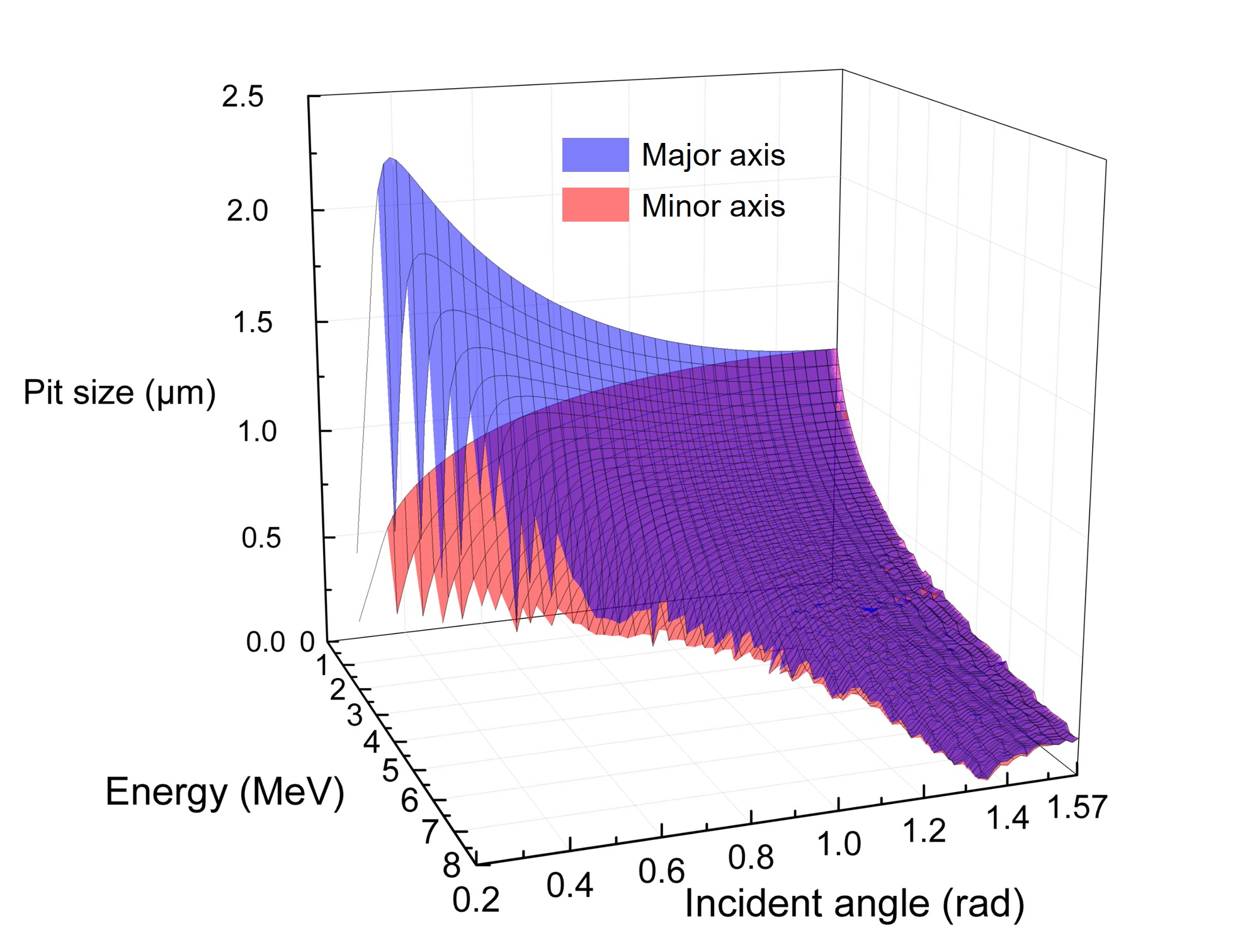}
    \caption{\label{fig:reconion}Proton energy and incident angle are reconstructed from the major and minor axes of the pit openings. This dataset can be extended to four dimensions by incorporating different etching durations, enabling an inverse mapping to retrieve the energy and angular distributions of the proton beam.}
\end{figure}

With the etch-rate ratio derived from the REL data using Eq.~\eqref{eq:ratioREL}, any initial ion property can be mapped to a corresponding pit structure under specified etching conditions. Figure~\ref{fig:reconion} shows the relationship between ion energy, incidence angle, and the resulting major and minor pit axes. For an arbitrary measured pit shape $(a_m, b_m)$, inverse mapping to the precomputed dataset yields candidate values $(E_m, \theta_m)$ that minimize the axis-length error within a user-defined tolerance.

Ambiguities arise when multiple ion characteristics produce similar pit openings, which can occur in measurements involving multiple ion species or low-energy ions. These ambiguities result from dimensionality reduction and are inherent to two-dimensional imaging analyses. To mitigate this limitation, additional constraints based on the experimental configuration, such as placing filters in front of the CR-39 detectors, can be introduced to restrict the solution space. Alternatively, extended etching can be used to further separate pit features, provided that a precise scanning and positioning system is integrated with the optical microscope and the detection algorithm.

\section{Conclusion\label{se:conc}}

In summary, we have developed and validated an automated framework for CR-39 pit detection and ion spectral reconstruction using laser-driven ion acceleration experiments. The method combines a Hough-transform-based detection scheme with vectorized least-squares ellipse fitting to achieve precise measurements of the major and minor axes of individual pit openings from two-dimensional micrographs. The implementation supports efficient GPU acceleration and maintains robustness under high-yield conditions with substantial pit overlap. By coupling the extracted geometries with a pit-formation model based on Monte Carlo simulations of restricted energy loss, the ion energy and incidence angle are reconstructed directly from the measured data. Calibration using multi-MeV proton beams generated via the target-normal sheath acceleration mechanism establishes a quantitative mapping between pit geometry and reconstructed ion parameters.

The proposed framework provides an efficient and scalable approach for high-precision ion-beam diagnostics, particularly in high-flux regimes where conventional analysis methods become unreliable. Future work will focus on extending the framework to multi-species discrimination, integrating automated microscopy with motorized stages for high-throughput analysis, and mitigating background artifacts from dust and surface defects through machine-learning-based techniques. The methodology can be extended to other solid-state track detectors and high-density particle imaging applications, offering a pathway toward automated, high-throughput diagnostics in laser-plasma experiments.

\section*{Supplementary Material}
The detailed mathematical derivations of the pit-formation model and the ellipse-fitting framework are provided in the supplementary material.

\section*{Acknowledgments}
This study was supported by the National Science and Technology Council in Taiwan (Grant Nos. NSTC 114-2119-M-008-007 and 113-2112-M-008-030-MY2).

\section*{Author Declarations}
\subsection*{Conflict of Interest}
The authors have no conflicts to disclose.

\subsection*{Author Contributions}
Wei Lin: Conceptualization (equal); Data curation (equal); Formal analysis (equal); Methodology (equal); Project administration (equal); Software (equal); Writing -- original draft (equal); Writing -- review \& editing (equal). Tzu-Yao Huang: Conceptualization (equal); Data curation (equal); Formal analysis (equal); Methodology (equal); Project administration (equal); Validation (equal). Shih-Fan Yang: Data curation (supporting); Formal analysis (supporting); Methodology (supporting); Project administration (supporting). Shih-Wei Wang: Data curation (supporting); Formal analysis (supporting); Project administration (supporting); Validation (supporting). Wu-Cheng Chiang: Data curation (supporting); Methodology (supporting); Project administration (supporting); Software (supporting). Kuo-Yu Hsiao: Data curation (supporting); Project administration (supporting). Chun-Han Chen: Data curation (supporting); Formal analysis (supporting); Methodology (supporting). Yan-Syu Liu: Data curation (supporting); Project administration (supporting). Jun-Yi Chen: Data curation (supporting); Project administration (supporting). Ya-Po Yang: Investigation (supporting); Project administration (supporting); Resources (supporting). Chen-Kang Huang: Conceptualization (supporting); Investigation (supporting); Project administration (supporting); Validation (supporting); Writing -- original draft (supporting). Hsu-hsin Chu: Conceptualization (supporting); Funding acquisition (supporting); Methodology (supporting); Project administration (supporting); Resources (supporting); Supervision (supporting). Chih-Hao Pai: Conceptualization (equal); Formal analysis (equal); Funding acquisition (equal); Investigation (equal); Methodology (equal); Project administration (equal); Resources (equal); Software (equal); Supervision (equal); Validation (equal); Visualization (equal); Writing -- original draft (equal); Writing -- review \& editing (equal).

\section*{Data Availability}
The data that support the findings of this study are available from the corresponding author upon reasonable request.

\bibliographystyle{aipnum4-2}
\bibliography{ref}

\end{document}